# Nuclear-spin-dependent coherent population trapping of single nitrogen vacancy centers in diamond

D. Andrew Golter, Khodadad N. Dinyari, and Hailin Wang

Department of Physics and Oregon Center for Optics

University of Oregon, Eugene, Oregon 97403, USA

Abstract

Coherent population trapping (CPT) provides a highly sensitive means for probing the energy level structure of an atomic system. For a nitrogen vacancy center in diamond, the CPT offers an alternative to the standard optically-detected magnetic resonance method for measuring the hyperfine structure of the electronic ground states. We show that the nuclear spin dependent CPT measures directly the hyperfine splitting of these states due to the $^{14}$N nuclear spin. The CPT spectral response obtained in the presence of a strong microwave field, resonant or nearly resonant with a ground state spin transition, maps out the dynamic Stark splitting induced by the coherent spin excitation.

When two atomic ground states are coupled to a common excited state via coherent optical fields, destructive quantum interference between the two transitions can lead to a trapping of the population in an optically "dark" superposition state[1, 2]. This effect, known as coherent population trapping (CPT), represents a useful tool for the optical control of quantum systems. An important quantum system where CPT can be realized is the electron spin states of nitrogen vacancy (NV) impurity centers in diamond. With ultra-long electron and nuclear spin decoherence times, single-shot spin detection, sub-nanosecond spin control, and efficient quantum state transfer between electron and nearby nuclear spins, diamond NV centers have emerged as a leading candidate for quantum information processing in a solid-state platform[3-9]. CPT can potentially provide a unique and effective mechanism for controlling and detecting electron and nuclear spins with a nanoscale spatial resolution[10, 11]. Such capabilities will be important for implementing scalable quantum logic gates in systems such as closely spaced chains or clusters of NV centers[12].

The first demonstrations of CPT via optical transitions in diamond NV centers relied on either strong magnetic fields or high levels of crystal strain to create a Λ-type three-level system via state mixing between the $m_s$=0 and $m_s=\pm 1$ electron spin states[13, 14]. This is in addition to coherent optical control of electron spins of NV centers[15, 16]. CPT was also observed within the microwave-driven electron spin transitions[17]. More recently, CPT was realized with the $m_s=\pm 1$ electron spin states as the two lower levels of a Λ-type three-level system, without resorting to either high magnetic field or high crystal strain[18]. For these latter experiments, a single optical field was used along with the application of a weak magnetic field to tune the two electron spin states in and out of the two-photon resonance associated with the CPT. This CPT experiment also demonstrated laser cooling of the nuclear spin environment of a given NV center through the underlying coherent optical interactions.

In this paper, we report an experimental study of CPT using the $m_s=\pm 1$ electron spin states in a single diamond NV center. Both optical and microwave fields are used to drive coherently the coupled electron-nuclear spin system. As shown in the earlier study, CPT is sensitive to the hyperfine coupling between the electron spin and nearby nuclear spins[18]. This nuclear-spin-dependent CPT process provides an effective means for probing the hyperfine energy level structure of the NV center and offers an alternative to the standard optically-detected magnetic resonance (ODMR) method for nuclear spin detection. The ODMR method depends on nuclear-spin-dependent, microwave-driven electron spin transitions[8], which is not needed for CPT, thus freeing up these microwave-driven transitions for other uses. For our experimental studies, a strong microwave field is applied resonantly or near resonantly to the transitions between the $m_s$=0 and $m_s=\pm 1$ electron spin states that are hyperfine coupled to a $^{14}$N nuclear spin with spin 1. We use the CPT generated by two coherent optical fields to measure the dynamic Stark effect induced by the microwave field. These studies demonstrate the effectiveness of CPT for spin detection, especially for probing coherent spin processes of the hyperfine-coupled electron-nuclear system in diamond.

The experimental studies reported here were carried out in a type IIa high purity diamond at low temperature (T ≈ 7 K) in a liquid-helium cryostat. Individual NV centers were excited and imaged using a standard confocal microscope setup, with the phonon-sideband fluorescence coupled into a single mode fiber and sent to an avalanche photodiode detector. To improve collection efficiency, we used a focused ion beam to etch a solid immersion lens (SIL) into the surface of the diamond and worked with a NV center located near the center of the SIL[19, 20]. A green 532 nm diode laser provided off-resonant excitation of the NV center, while a red 637 nm tunable ring dye laser provided resonant optical excitation. Acousto-optic modulators (AOMs) were used to generate the desired optical pulse sequence as well as the relative frequency detuning between two optical fields derived from the same laser. A thin (approximately 20 μm in diameter) bonding wire stretched over the surface of the diamond acted as an antenna for supplying microwave fields.

Figure 1a shows the Λ-type three-level configuration, with the $m_s=\pm 1$ electron spin states forming the two lower levels. The electronic ground state of the negatively charged NV center is a spin triplet, which is split by spin-spin interactions into $m_s$=0 and $m_s=\pm 1$ states[21, 22]. Spin conserving transitions between the ground state and six different excited states can be driven optically[21-24]. For the Λ-type three-level system shown in Fig. 1a, state $A_2$ serves as the upper state. In the absence of strain and with no external magnetic field, an electron in state $A_2$ decays radiatively to the $m_s$ =+1 and $m_s$=-1 states with equal probability. The three energy levels together form a nearly closed system, with only a small probability for an electron in state $A_2$ to decay nonradiatively into the $m_s$ =0 state via a metastable singlet state. This three-level system thus also provides a good model system for generating quantum entanglement between a photon and an electron spin[25].

To confirm which of the ground states is involved in a given optical transition, we drove Rabi oscillations between the $m_s$=0 state and the $m_s$=$\pm 1$ state using a microwave pulse. The microwave pulse was preceded by a green optical pulse, which initialized the NV center into the $m_s$=0 state. The NV center was then excited by an optical pulse resonant to a given optical transition. The inset of Fig. 1b shows the pulse sequence of the experiment, indicating the timing of the green, microwave, and resonant optical pulses as well as the fluorescence detection. Figure 1b shows the fluorescence from the NV center as a function of the microwave pulse duration, with the NV center resonantly excited through the $A_2$ transition (bottom trace) and the $E_y$ transition that couples to the $m_s$ =0 state (top trace). The phase of the Rabi oscillations observed verifies which ground state population was being measured.

A small, static magnetic field, supplied by a permanent magnet, was used in the CPT experiment to induce a Zeeman splitting of the $m_s$=$\pm 1$ states[26, 27]. In the limit of zero strain, state $A_2$ couples to the $m_s$ =+1 and $m_s$ =-1 states via σ- and σ+ polarized light, respectively. For non-zero strain, as was the case for the sample used in our experiment, the pure circular polarization selection rule no longer holds. The Zeeman spin splitting enables incident optical fields to couple to specific optical transitions of the Λ-type three-level system, even when the NV center is not in a strain-free environment. Figure 1c shows the photoluminescence excitation (PLE) spectra of the two individual $A_2$ transitions, with a Zeeman splitting of approximately 500MHz (smaller Zeeman splitting was used in the CPT experiments). For this experiment, a CW microwave field was applied to the transition between the $m_s$=0 state and either the $m_s$ =+1 or $m_s$ =-1 state, populating a given spin state for the respective optical transition. Note that resonant optical excitations can ionize the NV centers. This ionization can be reversed by off-resonant (green) excitation[16]. For this reason, the resonant optical field was also alternated with a green optical pulse. Since the green optical pulse can change the electronic environment of the NV center, the resulting spectral diffusion leads to a PLE linewidth (≈700 MHz) that is much greater than the intrinsic radiative linewidth[28, 29].

For the CPT experiment, two linear, perpendicularly polarized optical fields with equal optical power coupled simultaneously to the two respective $A_2$ transitions. The frequency of one of the fields was tuned, while the other was held fixed. The overall polarization of the two optical fields was adjusted with a half-wave plate until the fluorescence was maximized. Again, alternating resonant and off-resonant optical pulses were used. To counteract the optical pumping due to the off-resonant excitation, we also applied a weak CW microwave field resonant with a ground-state spin transition.

Figure 2a shows the fluorescence observed as a function of the detuning between the two optical fields. A pronounced dip occurs in the excitation spectrum when the detuning approaches the Zeeman splitting, indicating that the NV center is driven into a dark state, which in this case is a nearly equal superposition of the $m_s$ =+1 and $m_s$ =-1 states. The width of the CPT dip is determined by the optical Rabi frequency. The relatively high optical power (1μW) used for Fig. 2a leads to a dip linewidth (≈16 MHz) that far exceeds the hypefine splitting (2.2 MHz) due to the coupling between the electron spin and $^{14}N$ nuclear spin. In this limit, the CPT process is not affected by the hyperfine coupling process.

The CPT process becomes dependent on the nuclear spin states, when the optical Rabi frequency is small compared with the hyperfine splitting. Due to the hyperfine coupling between the electron spin and the $^{14}N$ nuclear spin in a NV center, the energy of the electron $m_s=\pm 1$ state depends on the orientation of the $^{14}N$ nuclear spin. Specifically, each electron spin state splits into three hyperfine states, corresponding to nuclear spin projection with $m_n$=-1, 0, +1 (see Fig. 2b)[26, 30]. CPT can be used to spectrally resolve the individual hyperfine-split states. For the experiment shown in Fig. 2c, the $m_s$=0 to $m_s$=+1 transition was driven with a strong microwave π-pulse (Rabi frequency ≈ 5 MHz), which populates the $m_s$ =+1 state regardless of the nuclear spin state. Two relatively weak optical fields with a combined incident optical power near 60 nW were used to couple to the two respective $A_2$ transitions (see Fig. 2d for the pulse sequence used for the experiment). As shown in Fig. 2c, three CPT dips with a frequency separation of 4.4 MHz were observed in the excitation spectrum obtained as a function of the detuning between the two optical fields. These dips represent nuclear spin dependent CPT and occur when the frequency difference between the two optical fields is equal to the Zeeman splitting plus twice the hyperfine splitting for a particular nuclear spin state. In this case, the two-photon resonance condition for the CPT is satisfied for the $m_s$=±1 hyperfine states with the same nuclear spin projection. The splitting between the CPT dips observed thus corresponds to twice the hyperfine splitting. The center of the CPT spectra depends on the Zeeman splitting and varies depending on the magnetic field strength. Note that in addition to the hyperfine splitting, a quadrupole interaction splits the $m_n$=±1 states from the $m_n$=0 state by 5 MHz for all electron spin states[30].

This splitting does not change the two-photon resonance condition for the CPT and thus does not affect the CPT spectral response.

We further confirmed the dependence of the CPT on the nuclear spin projection by combining the CPT experiment with a nuclear-spin-dependent electron spin flip. Figure 2e shows the result of the CPT experiment where the microwave π-pulse is relatively weak, with a Rabi frequency (≈0.8 MHz) small compared with the hyperfine splitting, and is resonant with one of the hyperfine $m_s$=1 states. In this case, only one CPT dip is visible in the excitation spectra. The spectral position of the dip depends on which hyperfine $m_s$=1 state was initially populated. CPT of a NV center can thus also serve as an effective detection for individual nuclear spin states.

CPT is also sensitive to the coherent spin dynamics of the coupled electron-nuclear spin system provided by the NV center. Figure 3 shows the results of a CPT experiment, in which a strong resonant or nearly resonant microwave field is present during the resonant optical excitation. Similar to the experiment shown in Fig. 2c, we initialized the system with a green pulse followed by a strong, resonant (now with the $m_s$=0 to $m_s$=-1 transition) microwave π pulse, which ensures that all the hyperfine states with $m_s$=-1 are initially populated. A second, strong microwave field was then applied to the $m_s$=0 to $m_s$=+1 transition. Experimentally, we measured the CPT spectral response at various fixed spectral positions for the second microwave field.

Figure 3a plots the PLE excitation spectra obtained as a function of the detuning between the two optical fields for three different powers of the second microwave field, with the frequency of this microwave field resonant with the $m_n$=0 hyperfine state (corresponding to the center CPT dip in Fig. 2c). The excitation spectra clearly show the dynamic Stark splitting of the center CPT dip (along with a less obvious shifting of the other two dips), corresponding to the Stark splitting of the hyperfine state with $m_n$=0. Figure 3b plots the excitation spectra obtained as a function of the detuning between the two optical fields at various fixed frequencies of the microwave field. In this case, the second microwave field was tuned across the transitions for the relevant hyperfine states, with the power held constant (Rabi frequency ≈ 4 MHz). As many as six CPT dips can now be observed in the excitation spectra.

In the above experiments, the spectral positions of the CPT dips are determined by the optical Stark effects induced by the strong second microwave field. In the dressed state picture, each hyperfine state is split into two dressed states, with the energy splitting determined by the Rabi frequency and detuning of the second microwave field[31]. These dynamic Stark effects can lead to six CPT dips in the excitation spectra. To describe the experimental results, we also plot in Fig. 3 the result of a calculation, in which we used the experimental parameters of Rabi frequency and detuning to determine the spectral positions of the CPT dips due to the dynamic Stark effects. The experimental excitation spectra were then numerically fit with six Lorentzians, with their center frequencies determined as discussed above and with the depths and linewidths of the CPT dips as adjustable parameters. Figure 3 shows an overall good agreement between the calculation and the experiment.

In summary, experimental studies of CPT of single NV centers in diamond demonstrate that this coherent nonlinear optical process can be highly effective in probing coherent as well as incoherent spin processes and in detecting nuclear spins hyperfine-coupled to an electron spin in a NV center. It was proposed recently that the use of CPT can in principle enable far-field optical spin manipulation and detection with spatial selectivity approaching a few tens of nanometers. It has been well established that CPT can be used for coherent spin manipulation through processes such as stimulated Raman adiabatic passage (STIRAP)[32]. The experimental studies presented here pave the way for further exploration using CPT for the nanoscale manipulation and detection of nuclear as well as electron spins, and in particular for probing and manipulating individual spins in a chain or cluster of closely spaced NV centers, in which scalable quantum logic gates can be implemented[12].

This work is supported by NSF under grant No. PHY-1005499 and by DARPA.

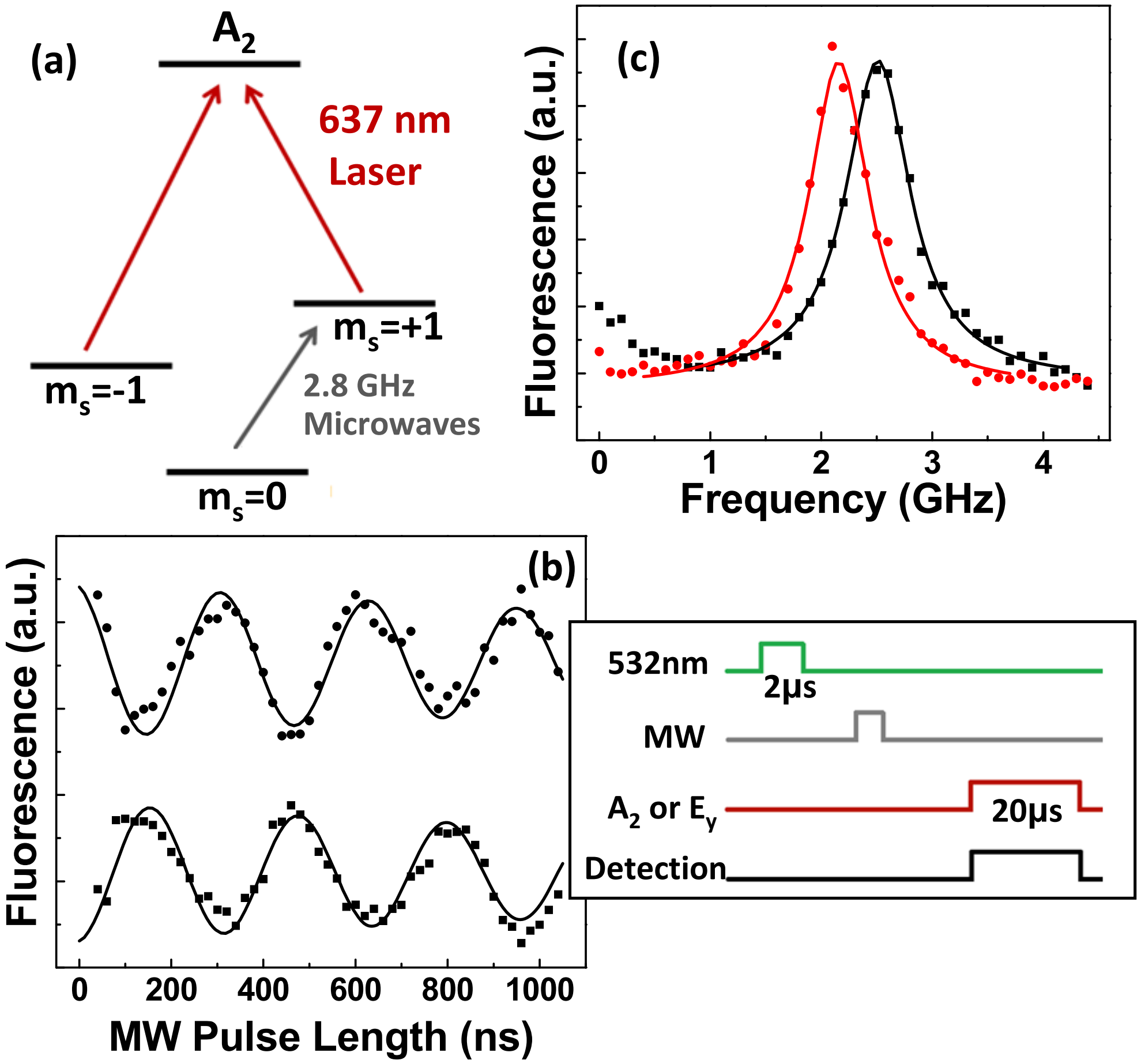

FIG. 1 (color online) (a) The Λ-type three-level system used for CPT, including a Zeeman splitting between the $m_s$=+1 and $m_s$=-1 states. (b) Rabi oscillations observed between the $m_s$=0 and $m_s$=±1 states (with no Zeeman splitting between the $m_s$=+1 and $m_s$=-1 states). Bottom trace: The resonant laser is tuned to the $m_s$=±1 to $A_2$ transition. Top trace: The resonant laser is tuned to the transition between the $m_s$=0 state and the $E_y$ excited state. Solid lines are a numerical fit to a damped periodic oscillation. Inset: The pulse sequence used. (c) PLE spectra of the two $A_2$ transitions, showing a Zeeman splitting of approximately 500 MHz. Solid lines are a Lorentzian fit.

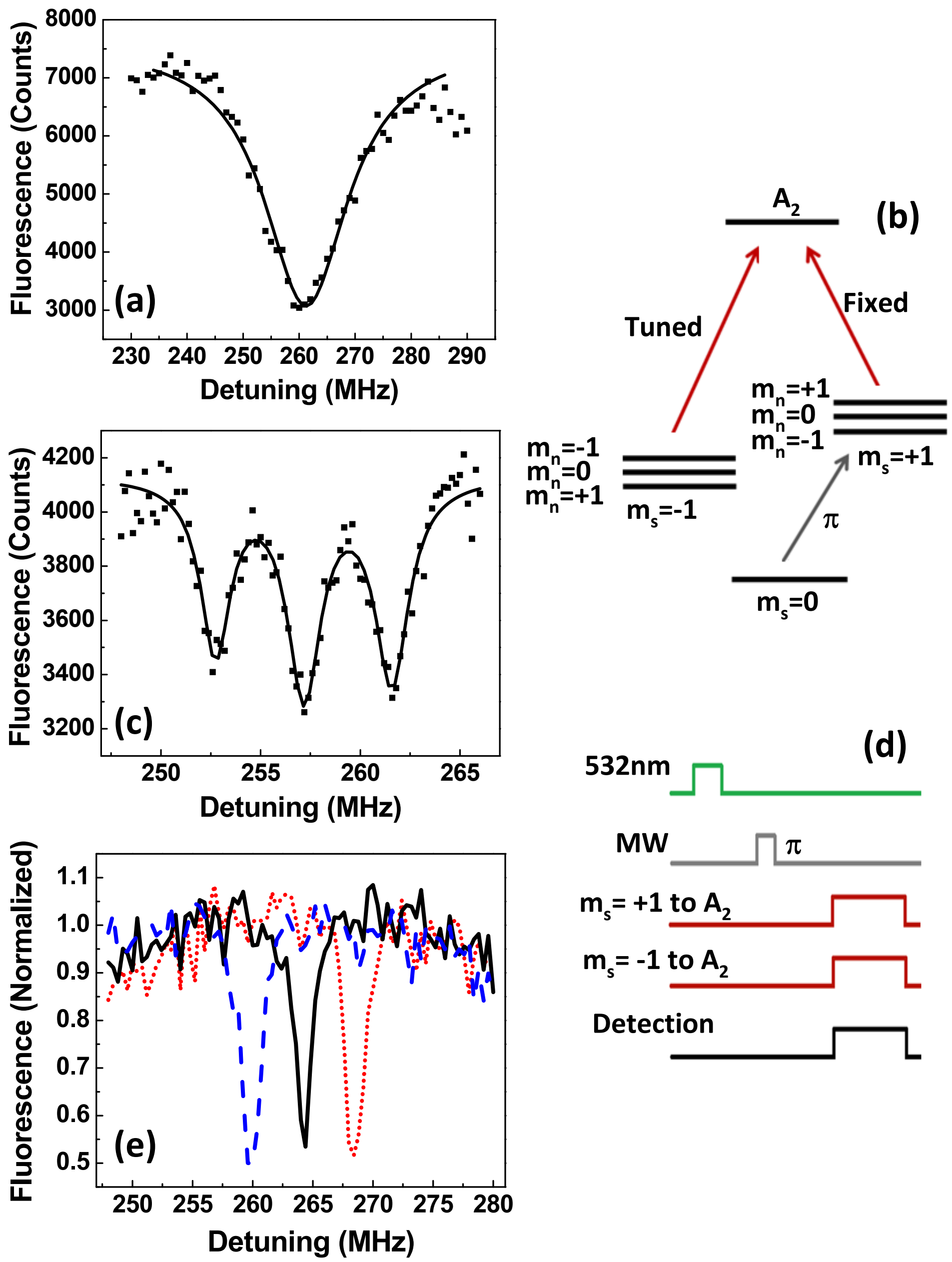

FIG. 2 (color online) (a) CPT spectral response with the Rabi frequency of the optical fields exceeding the hyperfine splitting. The solid line is a Lorentzian fit. (b) The energy level diagram with the hyperfine states labeled by their nuclear spin projection, $m_n$. (c) CPT spectral responses obtained with relatively weak optical fields and with a strong microwave π pulse populating all three hyperfine states with $m_s$=1. The solid line shows a fit to three Lorentzians. (d) Pulse sequence used for the nuclear spin dependent CPT. (e) CPT spectral response obtained with relatively weak optical fields and with a week microwave π pulse populating the $m_n$=-1 (blue or dashed), $m_n$=0 (black or solid), or $m_n$=+1 (red or dotted) state (with $m_s$=1).

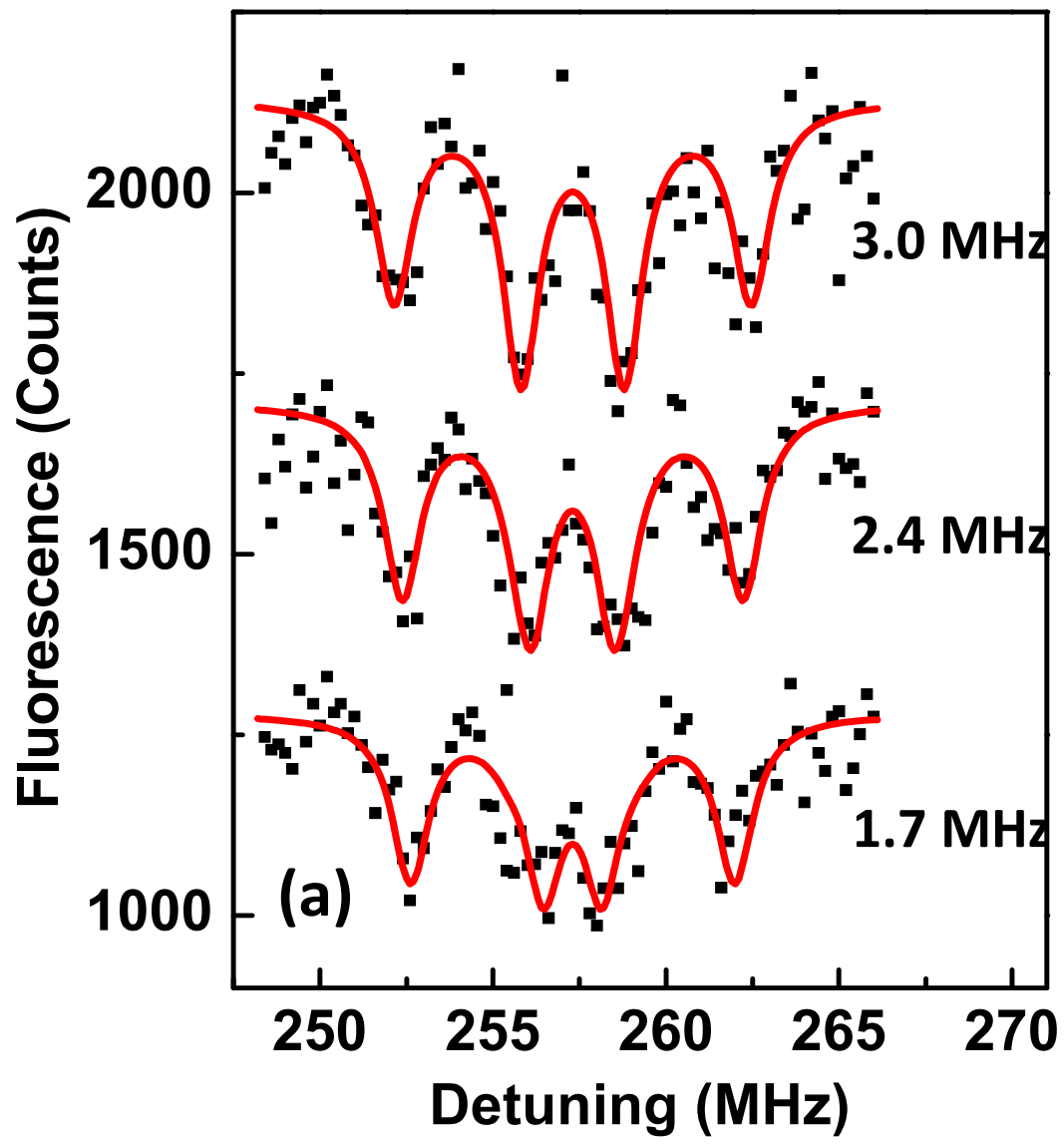

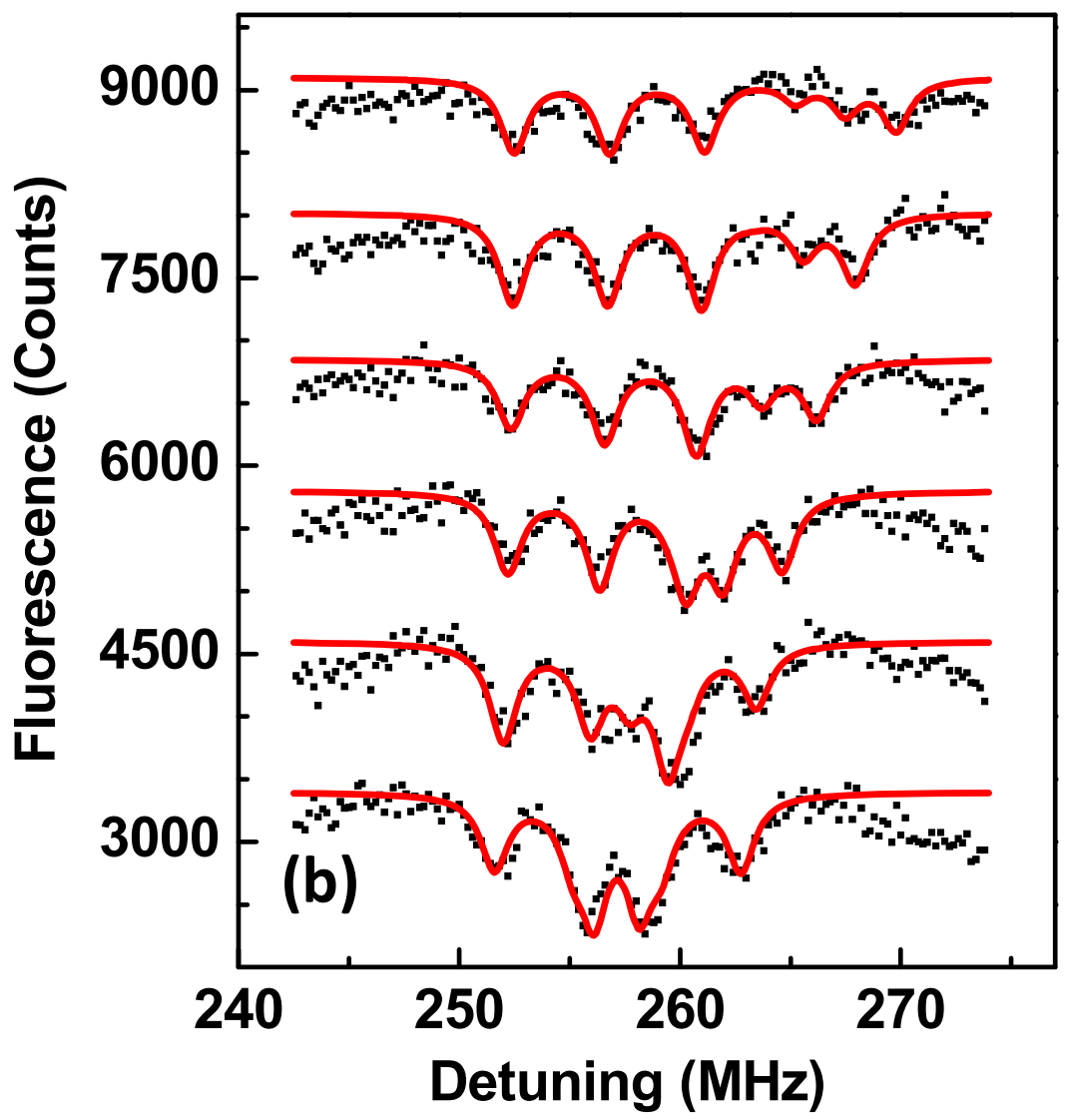

FIG. 3 (color online) CPT spectral responses (vertically offset) in the presence of a strong resonant or nearly resonant second microwave field. (a) Responses obtained with the microwave field resonant with the $m_n$=0 hyperfine state and with three different microwave powers. The traces are labeled with the microwave Rabi frequency. (b) Responses obtained with different microwave frequencies and with a fixed microwave power. For the top trace, the microwave frequency is 9.9 MHz above the spin transition for the $m_n$=0 state. For the succeeding traces, the frequency is shifted down in steps of 2 MHz. In both (a) and (b), solid lines show a fit to six Lorentzians, with the center frequencies determined theoretically based on the microwave power and frequency and with the widths and depths of the dips as adjustable parameters.